\documentclass{Interspeech2024}

\interspeechcameraready

\title{Quantifying the effect of speech pathology on automatic and human speaker verification}

\name[affiliation={1,2}]{Bence Mark}{Halpern}
\name[affiliation={3,4}]{Thomas}{Tienkamp}
\name[affiliation={1}]{Wen-Chin}{Huang}
\name[affiliation={1}]{Lester}{Phillip Violeta}
\name[affiliation={3}]{Teja}{Rebernik}
\name[affiliation={4}]{Sebastiaan}{de Visscher}
\name[affiliation={4}]{Max}{Witjes}
\name[affiliation={3}]{Martijn}{Wieling}
\name[affiliation={3}]{Defne}{Abur}
\name[affiliation={1}]{Tomoki}{Toda}

\address{
  $^1$Nagoya University, Japan 
  $^2$Netherlands Cancer Institute, The Netherlands \\
  $^3$University of Groningen, The Netherlands 
  $^4$University Medical Center Groningen, The Netherlands}
\email{halpern.bence.e8@f.mail.nagoya-u.ac.jp, tomoki@icts.nagoya-u.ac.jp}

\keywords{speaker verification, speech pathology, oral cancer}

\begin{document}

\maketitle

\begin{abstract}
This study investigates how surgical intervention for speech pathology (specifically, as a result of oral cancer surgery) impacts the performance of an automatic speaker verification (ASV) system. Using two recently collected Dutch datasets with parallel pre and post-surgery audio from the same speaker, NKI-OC-VC and SPOKE, we assess the extent to which speech pathology influences ASV performance, and whether objective/subjective measures of speech severity are correlated with the performance.
Finally, we carry out a perceptual study to compare judgements of ASV and human listeners. Our findings reveal that pathological speech negatively affects ASV performance, and the severity of the speech is negatively correlated with the performance. There is a moderate agreement in perceptual and objective scores of speaker similarity and severity, however, we could not clearly establish in the perceptual study, whether the same phenomenon also exists in human perception.

\end{abstract}

\section{Introduction}

Automatic speaker verification (ASV) has seen a large development  and is a crucial part of biometric systems. Recent studies, however, found that features used by ASV systems are useful for other novel applications. One example is predicting the severity of voice and articulation disorders \cite{quintas2022automatic, moro2020using}. Given that ASV systems are designed to distinguish individuals based on unique voice features, it is not surprising that ASV learns some pathology-related features. However, this realisation strongly hints that the effectiveness of ASV systems might be reduced with pathological speakers, and raises important questions. For example, is the severity of a speech disorder negatively correlated with the performance?

Another interesting application case is in the perceptual experiments of voice conversion (VC) \cite{wang2023duta, zhang2022sig}. In VC, an important aspect of performance evaluation is ensuring the converted voice's perceptual similarity to the target speaker's voice. This application case further raises the question: if the ASV system can be influenced by speech pathology, is human perception of speaker identity also influenced? If this phenomenon is not observable in human perception, ASV systems might not be an appropriate replacement for perceptual experiments in VC.

Despite these valid questions and the fact that there is a growing concern in the speech community about how models perform for atypical speech \cite{feng2021quantifying}, the impact of speech pathology on ASV remains largely unexplored. 
Previous research on the effect of pathological ASV has provided valuable insights but also left critical questions unanswered. For instance, the work of \cite{arasteh2023effect} showed that patients with conditions affecting the larynx face most
challenges with ASV, in contrast to articulation, and phonation disorders.
However \cite{arasteh2023effect} bases its analysis on cross-sectional data, which cannot exclude the difference between the speaker populations as a confound.  
While \cite{thienpondt2023behavioral, halpern2023improving} provided parallel data and reported similar results, it still remains unclear whether ASV speaker similarity scores align with those provided by human listeners.

Given these considerations, in this study, we evaluate the state-of-the-art (SotA) ECAPA-TDNN's \cite{desplanques2020ecapa} performance on parallel pathological and healthy\footnote{We refer to this speech as 'healthy' in the rest of the study, but it is important to note that the tumour's presence can already impede speech production in pre-operative speech.} data using two recently collected Dutch datasets: NKI-OC-VC \cite{halpern2023improving} and SPOKE \cite{Tienkamp2024ISSP_AAVS_ITOC}. Our aim is to provide a clearer understanding of how speech pathology affects ASV systems, identify their appropriateness for their current use cases (biometric, severity evaluation, perceptual), and
based on our results, show potential pathways for improving their robustness and accuracy for these use cases.

Our research questions are as follows:

\begin{enumerate}[label=\textbf{RQ\arabic*},noitemsep]
\item Is ASV performance (as measured by the equal error rate) reduced due to speech pathology (specifically, one caused by oral cancer surgery)?
\item If speech pathology affects ASV, is the severity of a speaker correlated with that speaker's ASV performance (as measured by the equal error rate)?
\item In general, what is the relationship between objective/subjective speaker similarity and objective/subjective severity difference of two utterances from the same speaker with different speech severity? Most importantly, if speech pathology affects ASV performance, is this phenomenon also observed in perceptual experiments?

\end{enumerate}

\section{Datasets}

\subsection{NKI-OC-VC}

The NKI-OC-VC dataset \cite{halpern2023improving} includes Dutch pathological speech from 6 oral cancer (OC) speakers (four male, two female) who had undergone a composite resection (COMANDO) surgery or comparable treatment for mostly advanced tongue tumours.

Data was collected from the participants at a maximum of three time points: before the surgery (T1), within a month after the surgery (T2), and approximately six months after surgery (T3). The recordings took place during scheduled speech therapy sessions. Participants were asked to read the Dutch text ``Jorinde en Joringel'' \cite{son01_eurospeech} consisting of 92 sentences during the recording session. The total duration of all speech recordings, across all speakers, was approximately 2.5 hours. One recording session (speaker/time point) lasted five minutes on average. In some cases, patients felt the experiment was difficult, in that case, we prematurely stopped the experiment.

The speech was recorded with a Roland R-09HR field recorder at 44.1 kHz sampling frequency and 24-bit depth. This was later downsampled to 16 kHz and quantized to 16-bit to match the requirement for the speaker embedding extractions. The dataset includes speech severity labels provided by Speech Language pathologists (SLPs) using a five-point Likert scale with 5 meaning healthy, and 1 meaning severe.

\subsection{SPOKE}
The SPOKE dataset collected by \cite{Tienkamp2024ISSP_AAVS_ITOC} was created to assess the acoustics and kinematics of individuals scheduled for surgery for early-stage lateral tongue tumours. The dataset contains Dutch speech data from a total of six participants (three male, three female). The speech was collected before the surgery (T1), and approximately six months after the surgery (T3). 
The speech was recorded in a sound-dampened booth \cite{wieling_spraaklab_2023}. Participants read 10 unique sentences during the recording session. The total duration of all speech recordings is approximately 20 minutes. One recording session (speaker/time point) lasted 1.5 minutes on average.

The speech was recorded using a Shure MX-153-T microphone with a sampling frequency of 22,050 Hz at 16-bit depth. After the recording, the audio was downsampled to 16 kHz and mixed to mono to match the requirements for the speaker embedding extraction.

\section{Methods}

\subsection{Objective speaker similarity (ASV) $sim_{obj}$}

For calculating the objective speaker similarity between two utterances, we extracted speaker embeddings using the ECAPA-TDNN \cite{desplanques2020ecapa} model from SpeechBrain \cite{ravanelli2021speechbrain}.\footnote{\url{https://huggingface.co/speechbrain/spkrec-ecapa-voxceleb}} 
In the trials, these embeddings were compared using the cosine similarity, which we term objective speaker similarity ($sim_{obj}$). All comparison trials were strictly on different utterances to avoid the impact of utterance similarity on the distributions. 

\subsection{Objective severity (P-ESTOI) $sev_{obj}$}
For calculating the objective severity of the speech, 
we used P-ESTOI \cite{janbakhshi2019pathological}. It involves creating a reference signal of intelligible speech from one or multiple healthy speaker(s), which is then time-aligned to the pathological speech signal using dynamic time warping (DTW). This alignment allows for a comparison of the signals and quantifies their divergence. P-ESTOI is demonstrated to yield high correlations with subjective intelligibility ratings, including OC speech \cite{halpern2023improving}. For each utterance, we calculate the P-ESTOI score by either taking the corresponding T1 utterance as a reference (for T2 and T3 utterances) or taking the average of all T1 utterances (for T1 utterances). 
\subsection{Subjective similarity and severity}

To obtain subjective scores, we carried out two subjective tests with 15 native Dutch raters on the SPOKE dataset. We could not carry out subjective speaker similarity experiments for the NKI-OC-VC due to restrictions on the data, and we used the subjective scores of SLPs. Participants had a mean age of 27.1 years (range = 24-30). For each test, we report the interrater correlation (\textit{ICC 2,k}). 

\textbf{Subjective similarity} ($sim_{sub}$) To obtain subjective similarity scores,  we carried out an experiment akin to the four-point scale voice similarity experiment often done in voice conversion (VC) studies, where two audio samples are presented to the listener side-by-side \cite{zhao2020voice}. However, for this study, we decided to adopt a 100 mm visual analogue scale (VAS) instead of the four-point scale that is common in VC studies. VAS is able to give a more nuanced range of uncertainty, which allows us to better understand speaker recognition phenomena. 

It is important that the listeners were unaware of the fact that there could be pathological speech in the dataset, specifically that there could be healthy and pathological speech from the same speaker. The instructions they received were therefore as follows (translation from Dutch): ``\textit{Please listen to the following two audio samples and rate them for speaker similarity. Please consider who is speaking according to the characteristics of the sound and then make a choice using the visual analogue scale that varies from ``Different (0\%)'' to ``Same (100\%)'' to rate the speaker similarity of the two audio samples. Please pick more extreme values if you are more confident in your choice. }''

The stimuli for the test included 55 comparisons, 25 same speaker same time point, 25 same-speaker different time point comparisons, and 5 different-speaker same time point comparisons. The stimuli were sampled based on the objective scores to make sure that we tested a broad range of comparisons. We did not use the scores given to different speakers in the analysis, but these were still needed to provide an appropriate range for the listening test. The average of the obtained similarity scores ($sim_{sub}$) were used in our further experiments.

\textbf{Subjective severity} ($sev_{sub}$) Following this test, listeners re-listened to the same stimuli, now rating them for speech severity using the same VAS method. We used the following instructions, which have been shown to provide scores comparable to expert listeners \cite{halpern2023automatic}: ``\textit{A "healthy" statement is 100\%, by which we mean that the pronunciation is easy to understand, you could write down the utterance, if asked, without any difficulty, and the speaker has clear articulation with normal speaking rate. A severely "pathological" utterance is 0\%, which means that the pronunciation is impossible to transcribe, the speaker has articulation problems, and the speaker may speak slower or faster than normal.}''

\section{Experiments}

\subsection{RQ1: Is automatic speaker verification impacted by the presence of pathology?}

For RQ1, we assessed the impact of speech pathology on ASV performance. We examined the effect of pathology on same-speaker and different-speaker $sim_{obj}$ distributions and measured ASV performance using the equal error rate (EER\%). 

To examine this effect, we break down $sim_{obj}$,  based on the speaker's identity and the timing of the recordings. For instance, comparisons between recordings from the same speaker at different times (e.g., T1 to T3) were labelled as ``Same T1-T3''. We compared these with a baseline (``Same T1-T1'' for same-speaker trials and ``Diff T1-T1'' for different-speaker trials), evaluating differences using a \textit{Wilcoxon rank-sum test}.

For EER calculations, we grouped these trials to represent various levels of difficulty. The ``T1-T1 EER'' scenario, the simplest, involves comparisons within the Same T1-T1 and Diff T1-T1 groups. The most challenging scenario, ``Any EER'', includes all same-speaker (``Same Any'') and different-speaker (``Diff Any'') comparisons. Additionally, for the NKI-OC-VC dataset, we examined two specific EERs: T12 EER, excluding comparisons with T3, and T13 EER, excluding comparisons with T2. For the SPOKE dataset, T12 EER cannot be calculated, and T13 EER is the same as ``Any EER''.

\subsection{RQ2: Speaker-severity vs speaker EER}

To investigate, whether severity is correlated with EER, we calculate "Any EER" for each speaker separately, and compare this with both the mean of objective severity scores $sev_{obj}$ for each T3 speaker, and the mean of the subjective severity scores obtained in $sev_{sub}$. 

\subsection{RQ3: General relationship between variables}

To understand the relationship between the subjective/objective speaker similarity and speech severity, we calculated a Pearson's correlations matrix between all the following properties of all the utterance pairs that we rated in the SPOKE dataset: the subjective speaker similarity ($sim_{sub}$), the objective speaker similarity ($sim_{obj}$), the objective speech severity difference ($\Delta sev_{obj}$) and the subjective speech severity difference ($\Delta sev_{sub}$). The latter two is calculated as the absolute difference of the two scores.

\section{Results}

\subsection{RQ1: Is ASV impacted by the presence of pathology?}

\begin{table}[ht]
\centering
\caption{Equal error rate (\%) when allowing for different time points in the same speaker comparison.}
\label{tab:eer_hter_values}
\resizebox{\columnwidth}{!}{%
\begin{tabular}{@{}ccccc@{}}
\toprule
\textbf{Dataset} & \textbf{T1-T1 EER (\%)} & \textbf{T12 EER (\%)} & \textbf{T13 EER (\%)} & \textbf{Any EER (\%)} \\
\midrule
NKI-OC-VC & 1.00 & 6.35 & 3.50 & 5.83 \\
SPOKE & 0 & -      & -      & 0.13 \\
\bottomrule
\end{tabular}
}
\end{table}

\begin{figure*}
    \centering
    \includegraphics[width=\textwidth]{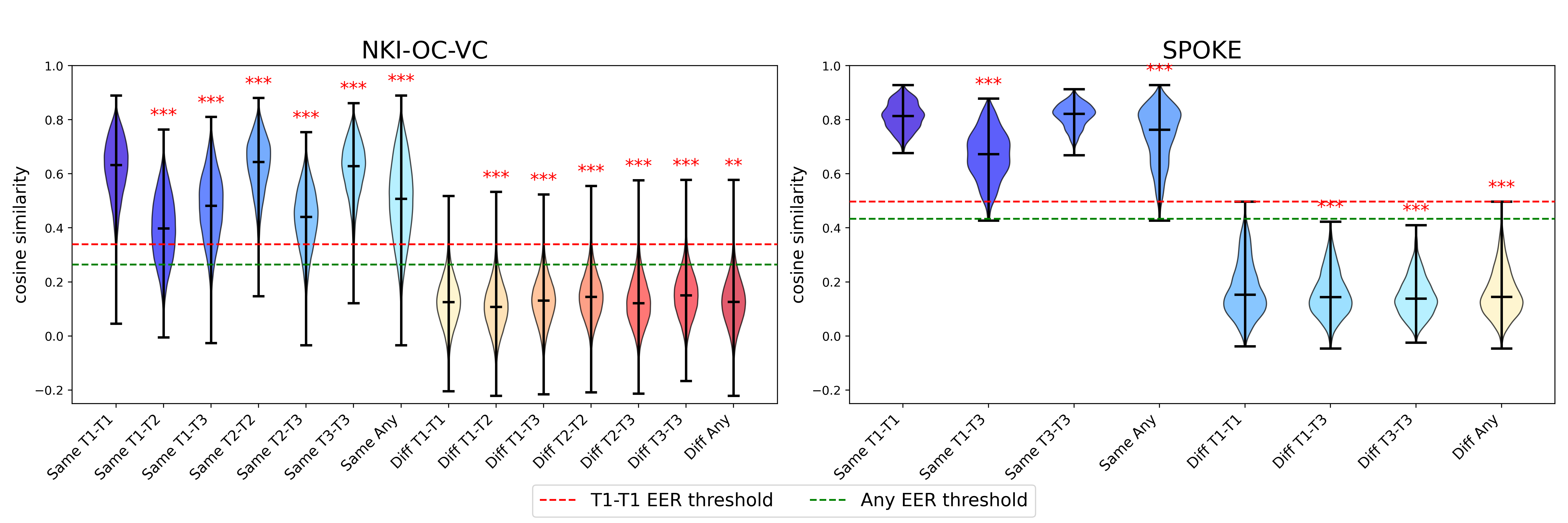}
    \caption{Comparison of cosine similarity distributions and EER thresholds for same-speaker and different-speaker trials across different time points.  Asterisks indicate the significance level of the correlations: (**) \(p < .01\), and (***) \(p < .001\).}
    \label{fig:eer_dist}
\end{figure*}

Figure \ref{fig:eer_dist} shows the objective similarity distributions for both datasets. In the NKI-OC-VC, the distributions for same-speaker trials show higher scores compared to different-speaker trials,  as expected. However, the presence of pathology shifts down the median of the distributions, most clearly in the Same T1-T2 comparisons, which suggests that the verification performance decreases most prominently at the T2 timepoint. All of the observed differences were significantly different from the  T1-T1 reference distributions. Note, however, that the different-speaker distributions' shift is marginal compared to the same-speaker distributions.
The SPOKE dataset shown a similar trend, with same-speaker trials yielding higher cosine similarity scores than different-speaker trials. The impact of pathology is also evident in this dataset, as Same T1-T3 and Same Any have a broader distribution and lower median value. However, no statistical difference was observed between Same T1-T1 and Same T3-T3.

The results in Table \ref{tab:eer_hter_values} summarise the impact of speech pathology in terms of EER results. For the NKI-OC-VC dataset, the EER increases when we allow comparisons with pathological speech. The highest EER rates are observed in the T12 case. This result is probably caused by speech directly after the surgery being more affected than six months after surgery due to scarring.

Conversely, the SPOKE dataset shows an absence of error rate (0\%) when the comparison is within the pre-surgery time point (T1-T1 EER), but a slight increase is observed when post-surgery utterances are included in the comparison (Any EER), reflecting a subtle influence of pathology on the system's performance. Again, the subtler increase can be accounted for the lower level of speech pathology in this dataset. We therefore conclude that speaker verification performance is impacted by speech pathology.

\subsection{RQ2: Speaker-severity vs speaker-EER}

\begin{figure}
    \centering
    \includegraphics[width=0.9\columnwidth]{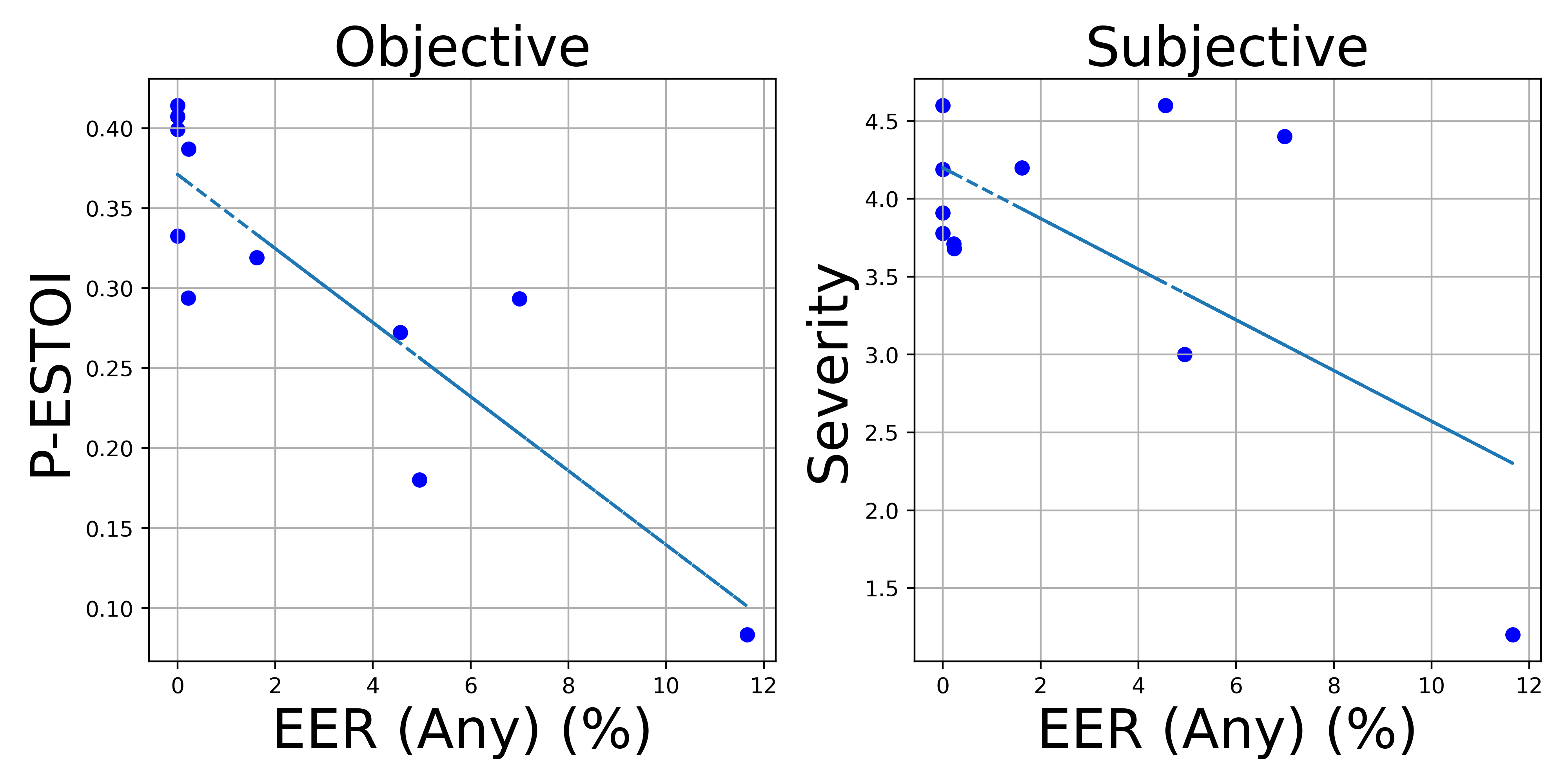}
    \caption{Objective and subjective speaker severity score in relation to EER performance.}
    \label{fig:eer_obj_sub}
\end{figure}

The analysis (shown in Figure \ref{fig:eer_obj_sub}) revealed a significant negative correlation between the EERs and objective severity scores ($r = -0.87, p < 0.05$), indicating that as the estimated severity of the speech pathology increases, a higher EER result can be expected. Furthermore, the study found a moderate positive correlation between the EERs and the subjective severity ($r = 0.65, p < 0.05$). However, the correlation is (much) weaker compared to the objective scores, which might be due to the non-ideal scenario of combining SLP and non-expert listeners scores in the experiment. Therefore, we can conclude that the severity is at least moderately predictive of the EER performance for a given speaker.

\subsection{RQ3: General relationship between variables}

\begin{table}[h]
\caption{Correlations between objective (\(\text{sim}_{obj}\)) and subjective (\(\text{sim}_{sub}\)) similarity measures, and objective (\(\Delta\text{sev}_{obj}\)) and subjective (\(\Delta\text{sev}_{sub}\)) severity difference measures. Asterisks indicate the significance level of the correlations: (*) \(p < .05\), (**) \(p < .01\), and (***) \(p < .001\).}
\label{tab:my-table}
\resizebox{\columnwidth}{!}{%
\begin{tabular}{@{}lcccc@{}}
\toprule
 & \multicolumn{1}{l}{\(\text{sim}_{obj}\)} & \multicolumn{1}{l}{\(\text{sim}_{sub}\)} & \multicolumn{1}{l}{\(\Delta\text{sev}_{obj}\)} & \multicolumn{1}{l}{\(\Delta\text{sev}_{sub}\)} \\ \midrule
\(\text{sim}_{obj}\)      & 1            & -           & -         & - \\
\(\text{sim}_{sub}\)      & 0.54 (***)   & 1           & -         & - \\
\(\Delta\text{sev}_{obj}\) & -0.28 (*)    & -0.15  & 1         & - \\
\(\Delta\text{sev}_{sub}\) & -0.43 (**)   & -0.16  & 0.40 (**) & 1 \\ \bottomrule
\end{tabular}%
}
\end{table}

\noindent

In the subjective tests, we observed moderate inter-rater reliability with both the similarity ($r=0.61$) and the severity ratings ($r=0.58$).

The correlation analysis, summarised in Table \ref{tab:my-table} shows a moderate positive correlation between the objective speaker similarity measures and subjective similarity ratings ($r=0.54$), suggesting that there is some level of agreement with the scores determined by the ASV and the human listeners.

The correlation between objective severity difference and subjective severity difference was not as high as expected ($r=0.40$), which may suggest that severity is not a property that is consistently expressed or perceived at the utterance level. While P-ESTOI has been successful in evaluating the intelligibility of pathological speech at the speaker level \cite{halpern2023improving, janbakhshi2019pathological}, its performance was less reliable at the utterance level \cite{halpern2023automatic}, indicating that the features that P-ESTOI captures may not translate directly to the perceptual cues listeners use to judge severity on the utterance level.

The relation between the speech severity difference and the speaker similarity is somewhat complicated. First, there is a lack of significant correlation between the subjective severity difference and the subjective similarity ($r=-0.16$), and also the objective severity difference and the subjective similarity ($r=-0.15$). However, there still seems to be a moderate correlation between the subjective severity difference and objective similarity ($r=-0.43$), which complicates the conclusion here. This inconsistency can partly be due to rater bias, as well as the limited set of samples used in our study compared to previous research which reported a higher interrater correlation \cite{halpern2023automatic}. 

Overall, we conclude that there is a moderate agreement on the scores between the human and machine listeners but the relationship between subjective similarity and severity cannot be clearly established.

\section{Discussion}

We find that the ASV systems are affected by speech pathology, and performance deterioration is correlated to the severity of speech, independent of whether this severity is measured via objective or subjective measures. We found moderate agreement between perceptual and objective scores. However, we could not directly observe that speech severity affects subjective speaker similarity scores.

These findings have several implications. Firstly, current biometric systems do not seem to be robust to variations in speech pathology. Therefore, it is advised that future research focuses on training techniques that increase the robustness against these variations. Examples of successful strategies in mitigating these issues from other speech technology tasks point towards data augmentation using voice conversion \cite{zhang2022mitigating} or other acoustic signal manipulations \cite{lin2023improving}. 

Secondly, it is clear from the experiment that current ASV models are also not well suited to replace perceptual experiments. Humans will naturally exhibit different responses due to their different biases, therefore this is not surprising. A notable example of this is familiar speaker recognition, which is known to activate different pathways compared to non-familiar speaker recognition \cite{van1985familiar}. 

Nevertheless, replacing perceptual experiments for voice conversion is an important goal, and several strategies could be used to finetune the ASV system towards that goal. E.g, instead of speaker labels, directly using similarity scores from raters might be a better (but arguably, more cost-intensive) approach to train systems that model perceptual aspects well. A similar strategy has been already used and shown to improve the quality of synthetic speech \cite{saito2021perceptual}. Another way for development to consider is using listener-dependent modelling, which has been shown to be an effective strategy in speech naturalness ratings \cite{huang2022ldnet}. Listener-dependent modelling is a straightforward way to model this bias in the systems.

Thirdly, it is interesting that ASV scores are competitive with P-ESTOI, as the method outperformed the P-ESTOI on the utterance level. The EER reduction was also highly predictive of the severity of the speaker's speech.

Finally, there are some limitations in our current analysis. First, our dataset comprises solely parallel data from low-to-mid speech severity speakers, where speaker embeddings are expected to be less impacted. Additionally, since the utterances used to compute the EER are from the same recording session, our estimation of the EER performance may be somewhat optimistic. Nonetheless, despite this potentially favourable scenario, we still observe a decline in performance, indicating that this limitation does not undermine the validity of our findings.

\section{Conclusion}

Our study showed that speech pathology negatively impacts automatic speaker verification (ASV) performance and performance reduction is negatively correlated with severity. We found moderate agreement between perceptual and objective scores, however, we could not establish the same phenomena in the perceptual experiments. The study highlights potential weaknesses in using these systems as biometrics and substituting perceptual experiments in voice conversion studies. We suggest strategies such as data augmentation, listener-dependent modelling, and similarity score training to alleviate these issues.

\section{Acknowledgements}
The data collection in the paper received ethical approval under the numbers IRBd20-159 (NKI-OC-VC), NL79242.042.21 (SPOKE), and ID95072353 (perceptual evaluation). The Department of Head and Neck Oncology and surgery of the Netherlands Cancer Institute receives a research grant from Atos Medical (H\"orby, Sweden), which contributes to the existing infrastructure for quality of life research. This work was further supported by the Research School of Behavioral and Cognitive Neurosciences of the University of Groningen.
This work is partly financed by the Dutch Research Council (NWO) under project number 019.232SG.011, and partly supported by JST CREST JPMJCR19A3, Japan.

\bibliographystyle{IEEEtran}
\bibliography{mybib}

\end{document}